\documentclass[letterpaper, 10 pt, conference]{ieeeconf} 
\IEEEoverridecommandlockouts
\usepackage{multirow}
\usepackage{cite}
\usepackage{amsmath,amssymb,amsfonts}
\usepackage{algorithmic}
\usepackage{graphicx}
\usepackage{textcomp}
\usepackage{tabularx}
\usepackage{amsmath}
\usepackage{dblfloatfix}
\usepackage{siunitx}
\usepackage{booktabs}
\usepackage{longtable}
\usepackage{float}
\usepackage{gensymb}
\usepackage{balance}
\usepackage{xcolor}

\title{\LARGE \bf
SEER: Semantic Enhancement and Emotional Reasoning Network for Multimodal Fake News Detection
}

\begin{document}
\author{Peican Zhu$^{1}$, Yubo Jing$^{1}$, Le Cheng$^{2}$, Bin Chen$^{3}$, Xiaodong Cui$^{4}$, Lianwei Wu$^{2}$, Keke Tang$^{5}$
\thanks{This work was supported in part by the National Natural Science Foundation of China (62472117), the Guangdong Basic and Applied Basic Research Foundation (2025A1515010157), the Science and Technology Projects in Guangzhou (2025A03J0137).
(Corresponding authors: Peican Zhu, Keke Tang)}
\thanks{$^{1}$Peican Zhu and Yubo Jing are with the School of Artificial Intelligence, Optics, and Electronics (iOPEN), Northwestern Polytechnical University, Xi’an, China (email:
ericcan@nwpu.edu.cn; jingyubo@mail.nwpu.edu.cn).}
\thanks{$^{2}$Le Cheng and Lianwei Wu are with the School of Computer Science, Northwestern Polytechnical University, Xi’an, China (email:
chengle@mail.nwpu.edu.cn; wlw@nwpu.edu.cn).}
\thanks{$^{3}$Bin Chen is with the Unit 93212 of People's Liberation Army of China, Dalian, China (email: 925190562@qq.com).}
\thanks{$^{4}$Xiaodong Cui is with the School of Marine Science and Technology, Northwestern Polytechnical University, Xi’an, China (email: xiaodong.cui@nwpu.edu.cn).}
\thanks{$^{5}$Keke Tang is with the Cyberspace Institute of
Advanced Technology, Guangzhou University, Guangzhou, China (email: tangbohutbh@gmail.com).}
}


\maketitle
\thispagestyle{empty}
\pagestyle{empty}

\begin{abstract}

Previous studies on multimodal fake news detection mainly focus on the alignment and integration of cross-modal features, as well as the application of text-image consistency.
However, they overlook the semantic enhancement effects of large multimodal models and pay little attention to the emotional features of news.
In addition, people find that fake news is more inclined to contain negative emotions than real ones.
Therefore, we propose a novel \textbf{S}emantic \textbf{E}nhancement and \textbf{E}motional \textbf{R}easoning (SEER) Network for multimodal fake news detection.
We generate summarized captions for image semantic understanding and utilize the products of large multimodal models for semantic enhancement. 
Inspired by the perceived relationship between news authenticity and emotional tendencies, we propose an expert emotional reasoning module that simulates real-life scenarios to optimize emotional features and infer the authenticity of news.
Extensive experiments on two real-world datasets demonstrate the superiority of our SEER over state-of-the-art baselines.

\end{abstract}

\section{INTRODUCTION}
With the rapid development of the internet, social media has become deeply embedded in people's daily lives and has become an indispensable part~\cite{lao2021rumor}.
Through social media, people can express their views, share insights, and post information anytime and anywhere.
With advantages such as strong timeliness, fast sharing speed, low production barriers, and convenient interaction, social media has significantly promoted information dissemination and exchange.
However, while facilitating information transmission, the rapid development of social media has also accelerated the production and spread of fake news, making social media platforms a breeding ground for misinformation~\cite{zhu2024general}, which poses serious threats to national security, economic development, and social stability~\cite{cheng2025efficient}.
Therefore, multimodal fake news detection becomes a problem that needs to be solved urgently.

\begin{figure}[t]
\setlength{\abovecaptionskip}{0cm}
\centering
\scalebox{0.75}
{\includegraphics[width=1\linewidth]{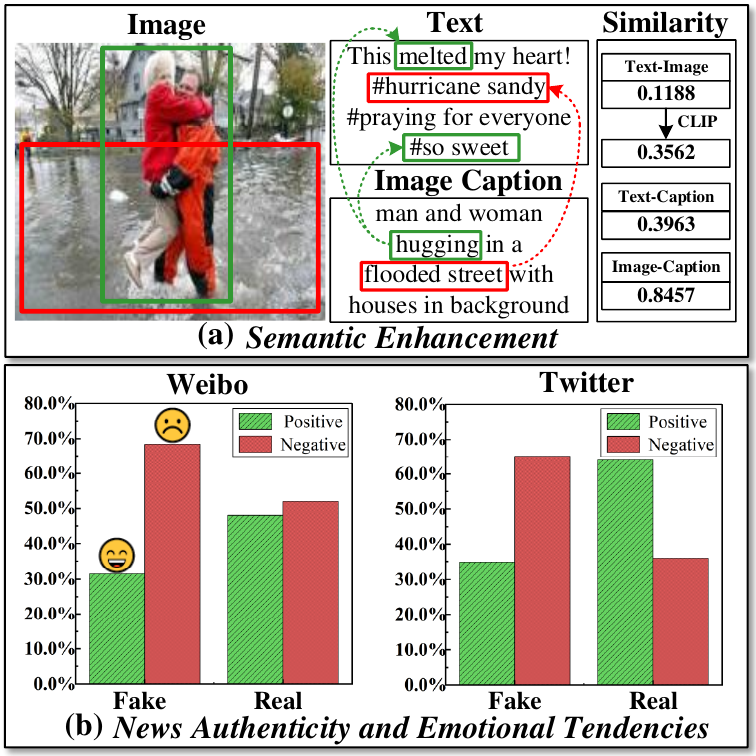}}
\caption{
An example of ignoring semantic enhancement and emotion analysis in previous studies.
Figure (a) demonstrates that the caption captures image semantics and enriches the content for news. 
Additionally, both the caption and CLIP can help improve cross-modal consistency.
Figure (b) shows the relationship between news authenticity and emotional tendencies, indicating that fake news is inclined to contain more negative emotions than real news.
}
\label{fig:statistics}
\end{figure}

Currently, multimodal fake news detection has gained tremendous development.
One area of research focuses on the enhancement, interaction, and fusion of multimodal features~\cite{Att-RNN,Spotfake,EM-FEND,zhu2024multimodal,HMCAN,zhu2025ken}. 
For example, MVNN~\cite{MVNN} extract and enhance features from images in both the frequency and pixel domains to help detect image manipulation in fake news.
KAN~\cite{KAN} extracts entities from both images and text and uses entity linking to obtain rich entity context, thereby enhancing the original multimodal features. 
Another class of research focuses on the application of consistency and similarity~\cite{chen2022cross,safe,FEND-CLIP,CCGN}.
For example, Safe~\cite{safe} considers that mismatched image-text content is more likely to be fake news, and thus uses cross-modal similarity to guide the detection process. 
CCGN~\cite{CCGN} employs a combination of contrastive learning, cross-modal consistency, and a graph neural network (GNN) to achieve excellent performance.

However, there are still shortcomings in previous works: 
1) \textbf{Insufficient semantic-level mining of images}. Existing methods~\cite{Att-RNN,MCAN,EM-FEND,KAN,KMGCN} only use pixel-level information or entity information of images.
While effective for detecting image manipulation, such as replacing people and background in Figure~\ref{fig:statistics}(a), these methods fail to deeply understand the actual semantics of image content and scenes. This leads to poor text-image alignment and fusion performance.
Therefore, we generate captions for images to enhance semantic understanding, thereby improving cross-modal alignment.
As in Figure~\ref{fig:statistics}(a), the “hugging” provides emotional context for the terms “melted” and “so sweet” in text. Additionally, “flooded street” reflects the actual situation depicted in the image and provides evidence for the credibility of “hurricane sandy” in text. 
By calculating and comparing similarity metrics, we further demonstrate the effectiveness of large multimodal models in narrowing the semantic gap.
2) \textbf{Limited use of emotion-level features}. Existing methods~\cite{MCAN,HMCAN,hua2023multimodal,yin2024gamc,CCGN} rarely focus on emotional features and overlook the differences between news articles with varying emotional tendencies.
Since the publisher creates fake news with malicious intent, fake news implicitly conveys negative emotions.
We conduct emotion analysis on two datasets.
As in Figure~\ref{fig:statistics}(b), there is a significant difference in  emotional tendencies between real and fake news on the Weibo and Twitter datasets, indicating that fake news is more likely to exhibit negative emotions.
Therefore, we can leverage the relationship between emotional tendencies and authenticity to optimize the model and improve detection accuracy.

In this paper, we propose a novel \textbf{S}emantic \textbf{E}nhancement and \textbf{E}motional \textbf{R}easoning (SEER) Network for multimodal fake news detection. We use the BLIP-2~\cite{BLIP} to generate the highly aligned image captions, which can be considered as descriptive information about the images for semantic understanding and knowledge expansion. In addition, we use the products of large multimodal models for semantic enhancement of unimodal and multimodal features. 
Inspired by people's perceived relationship between news authenticity and emotional tendencies, we construct the expert emotional reasoning module. We design multiple experts to conduct comprehensive evaluations of emotions from different modalities. Subsequently, we adjust contributions of emotions from different modalities to obtain final emotional scores and features of news. Finally, the emotional scores are used to predict news and optimize emotion features based on the perceived relationship between news authenticity and emotional tendencies.
Extensive experiments confirm the effectiveness and superiority of our SEER.

Our contributions can be summarized as follows:

1) We generate summarized captions for image semantic understanding and design structures to employ the products of large multimodal models for semantic enhancement.

2) We propose the expert emotional reasoning module, which considers the varying contributions of emotions from different modalities and uses the perceived relationship between news authenticity and emotional tendencies for fake news detection.

3) Extensive experiments show that our proposed SEER significantly outperforms several state-of-the-art baselines on two competitive datasets.

\section{Related Work}

Multimodal fake news detection has received increasing attention from scholars due to real-world demands.

\textbf{Interaction Fusion.}
Att-RNN~\cite{Att-RNN} used social comments to enrich textual features, while SpotFake~\cite{Spotfake} combined BERT~\cite{BERT} and VGG-19~\cite{VGG} for multimodal feature extraction.
EANN~\cite{EANN} introduced event detection task to extract event-invariant features.
MVAE~\cite{MVAE} utilized a variational autoencoder to reconstruct data and introduced adversarial training to boost feature representation.
Further studies~\cite{EM-FEND,KAN,KMGCN} enhanced feature aggregation by using visual and textual entities.
For example, KMGCN~\cite{KMGCN} built a GNN over images, text, entities, and knowledge nodes to enhance semantics.
Additionally, MCAN~\cite{MCAN} and HMCAN~\cite{HMCAN} employed co-attention~\cite{co-attention} to strengthen inter-modal interaction and fusion.
However, they overly focus on pixel-level information and entity information in images but fail to understand scenes and semantics. 
The insufficient alignment of cross-modal features results in poor fusion effectiveness.

\textbf{Consistent Alignment.}
Considering that news with mismatched image-text content is likely to be fake, consistency and discrepancy have been widely applied.
SAFE~\cite{safe} measured cross-modal similarity between textual and visual features.
CAFE~\cite{chen2022cross} proposed cross-modal ambiguity to balance unimodal and multimodal features, 
while works like FEND-CLIP~\cite{FEND-CLIP} and MMFN~\cite{MMFN} utilized CLIP~\cite{CLIP} to further learn the semantic consistency across different modalities.
However, many real news also suffer from mismatches between visual and textual content, so relying solely on consistency is unreliable.
Recently, CMMTN~\cite{CMMTN} utilized a masked attention to capture both intra- and inter-modal dependencies, while adopting curriculum-guided PU learning to boost detection effectiveness. 
Meanwhile, CCGN~\cite{CCGN} used contrastive learning to refine visual-textual alignment and combined cross-modal consistency with GNN to improve performance. 
However, existing works rarely considered emotional characteristics, overlooking the perceived relationship between news authenticity and emotional tendencies.

\section{Methodology}
\begin{figure*}[t]
\centering
\setlength{\abovecaptionskip}{0cm}
\scalebox{0.99}{\includegraphics[width=18cm]{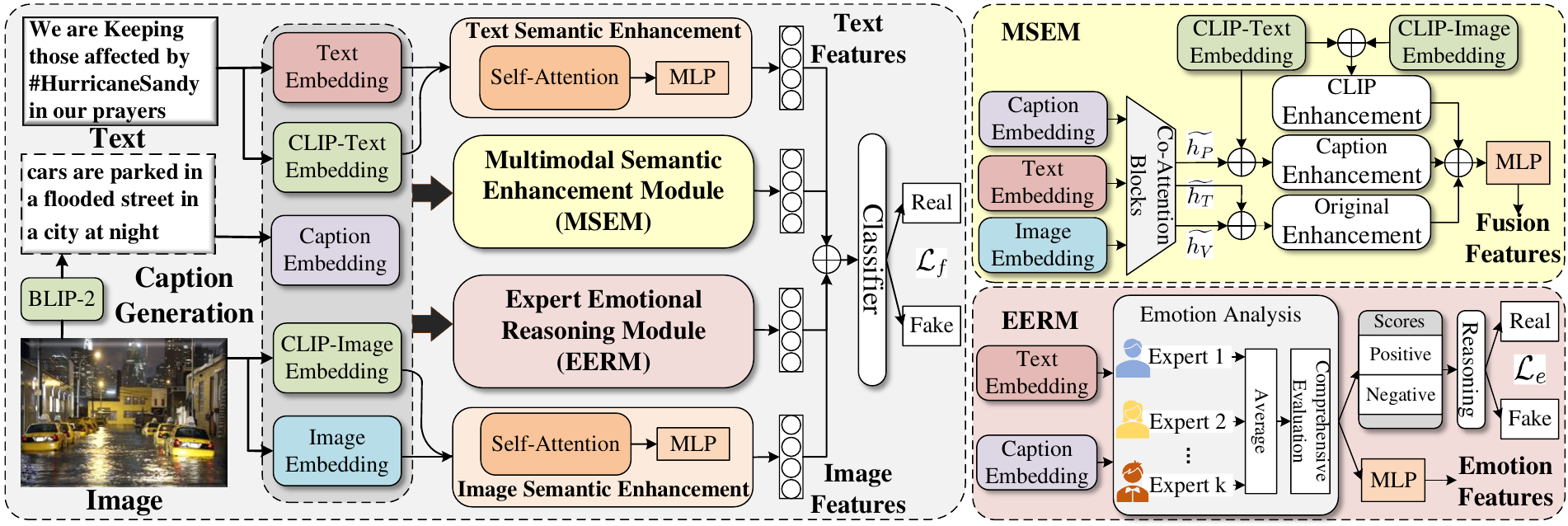}}
\caption{Illustration of the proposed SEER. It contains of four components: Feature Representations, Multimodal Semantic Enhancement, Expert Emotional Reasoning and Fake News Detector.}
\label{fig:method}
\end{figure*}

The overall structure is illustrated in Figure~\ref{fig:method}. 
We first extract feature embeddings from different modalities. 
In addition, we generate summarized image captions for semantic understanding and knowledge expansion. 
After that, inter-modal interaction is performed through co-attention blocks and three types of multimodal features are proposed. 
Intra-modal interaction is performed through self-attention and CLIP is used for unimodal semantic enhancement.
Then, we conduct evaluations of emotions in news by employing multiple experts and adjusting the importance of emotions from different modalities.
Subsequently, emotion features are optimized by leveraging the perceived relationship between news authenticity and emotional tendencies.
Finally, all features are integrated and fed into detector for classification.

\subsection{Feature Representations}

In this work, we focus on text and images. We assume that the $i$-th piece of social media news can be defined as $N_i=\{T_i, V_i\}$.
\noindent{\textbf{Text Embedding.}}
We use the pre-trained BERT~\cite{BERT} as the encoder for the news text $T$. The text embedding of the $i$-th news is denoted as $T_i=\{w_1,w_2,\cdots,w_m\}$, where $m$ is the length of the text.
\noindent{\textbf{Image Embedding.}}
We use Swin-T~\cite{Swin-T} as the encoder for news images $V$, the image embedding of the $i$-th news is denoted as $V_i=\{v_1,v_2,\cdots,v_n\}$, where $n$ is the number of regions in each image.
\noindent{\textbf{CLIP Enhanced Embedding.}}
We utilize the CLIP model~\cite{CLIP} to obtain the highly aligned text embedding $w_i^c$ and image embedding $v_i^c$ of the $i$-th news.
\noindent{\textbf{Caption Embedding.}}
We use BLIP-2~\cite{BLIP} to generate captions $P$ as descriptive information for semantic understanding of the images.
After that, we employ BERT to obtain the caption embedding $P_i=\{p_1,p_2,\cdots,p_z\}$ of the $i$-th news, where $z$ is the length of the caption.

\subsection{Multimodal Semantic Enhancement}
\noindent{\textbf{Inter-modal Interaction via Co-Attention.}}
Due to the cross-modal gap among text $T_i$, image $V_i$, and caption $P_i$, we employ three co-attention blocks, each consisting of two Transformer~\cite{BERT} encoders, to facilitate semantic interaction and enhancement across modalities.
The following is illustrated with text $T_i$ and image $V_i$ as inputs to a Transformer encoder. 
The attention distribution on $T_i$ is obtained through the similarity of the dot product between $V_i$ and $T_i$. 
Subsequently, textual information that is extremely relevant to the image can be extracted.
In order to comprehensively learn the relationship between $V_i$ and $T_i$, multihead attention mechanism performs $m$ parallel operations:
\begin{equation}
\begin{gathered}
h_i=softmax(\frac{(V_iW_j^q)\cdot(T_iW_j^k)^T}{\sqrt{d_h}})\cdot(T_iW_j^v)\\
MultiHead(V_i,T_i,T_i)=[h_1;h_2;...;h_m]\cdot W^o
\end{gathered}
\end{equation}
where $V_i$, $T_i$, $T_i$ are query, key, and value matrix, respectively. $W_j^q$, $W_j^k$, $W_j^v$ are the projection matrices for the $j$-th head, the symbol ; denotes the concatenation of vectors and $W^o$
is a trainable parameter.
Then, the important information obtained from the text is fused with the original image $V_i$ for learning and enhancement. Finally, we obtain the image representation $\tilde{h}_{t\to v}$ enhanced by text:
\begin{equation}
\begin{gathered}
h^\prime=Norm(V_i+MultiHead(V_i,T_i,T_i))\\
\tilde{h}_{t\to v}=Norm(h'+FFN(h'))
\end{gathered}
\end{equation}
where $Norm(\cdot)$ and $FFN(\cdot)$ are the normalization method and feed forward network.
In the same co-attention block, it is also possible to learn that the text representation $\tilde{h}_{v\to t}$ enhanced by image. Similarly, we can obtain $\tilde{h}_{p\to t}$, $\tilde{h}_{t\to p}$, $\tilde{h}_{p\to v}$, and $\tilde{h}_{v\to p}$ by the other two co-attention blocks. We
apply an averaging operation on the same modality to obtain the final semantically enhanced representations of the text, image, and caption, i.e., $\widetilde{h_T}$, $\widetilde{h_V}$, and $\widetilde{h_P}$, respectively.

\noindent{\textbf{Multimodal Feature Fusion.}}
Due to the existing gap between the products of large multimodal models and the original modalities, we propose three types of multimodal features. These features are both obtained by fusing and learning from the text-related and image-related semantic information. They are original multimodal features $\mathcal{M}_1=\sigma_1(concat(\widetilde{h_T},\widetilde{h_V}))$, CLIP-enhanced multimodal features $\mathcal{M}_2=\sigma_2(concat(w_i^c,v_i^c))$, and multimodal features $\mathcal{M}_3=\sigma_3(concat(w_i^c,\widetilde{h_P}))$ learned through the interaction of BLIP-2 and CLIP. $\sigma_i(\cdot)$ indicates feed forward network for mapping features to same space.
Finally, three types of features are passed to MLP, obtaining final fusion features $\mathcal{M}_{ter}$ after inter-modal interaction and semantic enhancement. 
In addition, we introduce the text-image similarity $\theta=\cos\left(w_i^c,v_i^c\right)$ to adjust relationships between unimodal and multimodal features:
\begin{equation}
\mathcal{M}_{ter}=\theta\cdot MLP_M(concat(\mathcal{M}_1,\mathcal{M}_2,\mathcal{M}_3))
\end{equation}

\noindent{\textbf{Unimodal Semantic Enhancement.}}
In order to explore global dependencies within each modality, 
we conduct intra-modal interaction using self-attention:
\begin{equation}
\tilde{w}=SelfAttention(T_i), \tilde{v}=SelfAttention(V_i)
\end{equation}
where $\tilde{w}$ and $\tilde{v}$ represent the textual and visual embeddings after intra-modal interaction.
We feed the enhanced features of CLIP with the features of intra-modal interaction into MLP for fusion and learning to achieve semantic enhancement of unimodal features. Finally, the unimodal features $T_{tra}$ and $V_{tra}$ for the text and image are obtained:
\begin{equation}
\begin{gathered}
T_{tra}=MLP_T(concat(w_i^c,\tilde{w}))\\
V_{tra}=MLP_V(concat(v_i^c,\tilde{v}))
\end{gathered}
\end{equation}

\subsection{Expert Emotional Reasoning}
\noindent{\textbf{Emotion Analysis.}}
We analyze the overall emotional tendencies of news by using both text $T_i$ and caption $P_i$ as inputs, which include both positive and negative ones. We prevent overly arbitrary assessments by employing $k$ experts to evaluate the emotional tone of each modality. For example, we feed text $T_i$ to an expert. 
First, a Bi-LSTM layer is used to learn the contextual emotional semantics. After that, a multihead self-attention is applied to learn the more important parts of the emotion features.
Finally, the textual emotion features $e_k^T$ are fed into MLP, and the $k$-th expert's emotional evaluation score $g_{e,k}^T$ for the text $T_i$ is obtained as:
\begin{equation}
\begin{gathered}
e_i^T=[ {\overset{\rightarrow}{e}}_i^T;{\overset{\leftarrow}{e}}_i^T]=BiLSTM(T_i)\\
e_k^T=SelfAttention(e_i^T)\\
g_{e,k}^T=softmax(W_Te_k^T+b_T)
\end{gathered}
\end{equation}
where the initial emotion features $e_i^T$ are obtained by concatenating the hidden layer outputs  ${\overset{\rightarrow}{e}}_i^T$ and ${\overset{\leftarrow}{e}}_i^T$, $W_T$ and $b_T$ are trainable parameters.
Similarly, we can obtain the emotion features $e_k^P$ and evaluation score $g_{e,k}^P$ of the caption $P_i$ from the $k$-th expert. 
After that, we perform the mean and normalisation operations on $k$ experts, obtaining the emotional scores $g_e^T$ and $g_e^P$, as well as the emotion features $e^T$ and $e^P$ for the text and caption, respectively. 
Due to the varying contributions of emotions from different modalities, we introduce the parameter $\lambda$ to adjust the emotional importance of different modalities.
Therefore, we obtain the overall emotional evaluation score $g_e\in(0,1)$, where a larger score indicates the news is more inclined to be positive.
In addition, we feed the adjusted emotion features from different modalities into MLP to obtain the final emotion feature $E$ of the news:
\begin{equation}
\begin{gathered}
g_e=\lambda g_e^T+(1-\lambda)g_e^P\\
E=MLP_E(concat(\lambda e^T,(1-\lambda)e^P))
\end{gathered}
\end{equation}

\noindent{\textbf{Emotional Reasoning.}}
We utilize the perceived relationship between news authenticity and emotional tendencies to predict the classification labels of news based on emotional scores, aiming to optimize emotion features. 
Initially, news can be considered equally likely to be real or fake before detection. Next, we introduce parameters $\alpha$ and $\beta$, which represent the probability of a positive emotional tendency in real and fake news, respectively.
According to Bayes' theorem, we can infer that the probability of positive emotional news being real is $\frac\alpha{\alpha+\beta}$, while the probability of negative emotional news being real is $\frac{1-\alpha}{2-\alpha-\beta}$.
Since $g_e$ represents the evaluation score for the positive emotion of the $i$-th news, we can ultimately obtain the predicted label $\hat{y}_i^{emo}$ for the news' authenticity as follows:
\begin{equation}
\hat{y}_i^{emo}=\frac\alpha{\alpha+\beta}g_e+\frac{1-\alpha}{2-\alpha-\beta}(1-g_e)
\end{equation}

We optimize emotion features for learning the perceived relationship between news authenticity and emotional tendencies, the emotional reasoning loss, i.e., $\mathcal{L}_{e}$, is defined as:
\begin{equation}
\mathcal{L}_{e}=\sum-[y_ilog(\hat{y}_i^{emo})+(1-y_i)log(1-\hat{y}_i^{emo})]
\end{equation}

\subsection{Fake News Detector}
We have obtained all the features for classification, including the text features ($T_{tra}$), image features ($V_{tra}$), emotion features ($E$) and fusion features ($\mathcal{M}_{ter}$). We concatenate the above features to form the final representation, i.e.,
$\mathcal{M}_{all}=\begin{bmatrix}T_{tra};V_{tra};E;\mathcal{M}_{ter}\end{bmatrix}$, which is passed through a fully-connected block to obtain the predicted probability of the $i$-th news. Then, cross entropy loss is employed as the rumor classification loss, i.e., $\mathcal{L}_{f}$, which is provided as:
\begin{equation}
\begin{gathered}
\hat{y}_i^{fnd}=softmax(W_f\mathcal{M}_{all}+b_f)\\
\mathcal{L}_{f}=\sum-[y_ilog(\hat{y}_i^{fnd})+(1-y_i)log(1-\hat{y}_i^{fnd})]
\end{gathered}
\end{equation}
where $W_f$ and $b_f$ are trainable parameters, $y_i$ indicates the ground truth label of the $i$-th news.
The overall loss for our SEER consists of emotional reasoning loss and rumor classification loss:
\begin{equation}
\mathcal{L}=\mathcal{L}_{e}+\mathcal{L}_{f}
\end{equation}

\section{Experiments}

\begin{table*}
\setlength{\abovecaptionskip}{0cm} 
\caption{Performance comparison of considered baselines and our SEER on two datasets.}
\centering
\begin{center}
\resizebox{0.70\textwidth}{!}{
\begin{tabular}{ccccccccc}
\hline
\multirow{2}{*}{Dataset} & \multirow{2}{*}{Methods} & \multirow{2}{*}{Accuracy} & \multicolumn{3}{c}{Fake News} & \multicolumn{3}{c}{Real News} \\ \cline{4-9} 
 &  &  & Precision & Recall & F1-score & Precision & Recall & F1-score \\ \hline
\multirow{11}{*}{Weibo} 
 & Att\_RNN~\cite{Att-RNN} & 0.772 & 0.854 & 0.656 & 0.742 & 0.720 & 0.889 & 0.795 \\
 & EANN~\cite{EANN} & 0.782 & 0.827 & 0.697 & 0.756 & 0.752 & 0.863 & 0.804 \\
 & MVAE~\cite{MVAE} & 0.824 & 0.854 & 0.769 & 0.809 & 0.802 & 0.875 & 0.837 \\
 & SpotFake~\cite{Spotfake} & 0.869 & 0.877 & 0.859 & 0.868 & 0.861 & 0.879 & 0.870 \\
 & SAFE~\cite{safe} & 0.816 & 0.818 & 0.815 & 0.817 & 0.816 & 0.818 & 0.817 \\
  & CAFE~\cite{chen2022cross} & 0.840 & 0.855 & 0.830 & 0.842 & 0.825 & 0.851 & 0.837 \\
 & MCAN~\cite{MCAN} & 0.899 & 0.913 & 0.889 & 0.901 & 0.884 & 0.909 & 0.897 \\
 & HMCAN~\cite{HMCAN} & 0.885 & 0.920 & 0.845 & 0.881 & 0.856 & 0.926 & 0.890 \\
 & CMMTN\cite{CMMTN} & 0.889 & 0.886 & 0.893 & 0.889 & 0.892 & 0.885 & 0.893 \\
 & CCGN~\cite{CCGN} & 0.908 & 0.922 & 0.894 & 0.908 & 0.894 & 0.922 & 0.908 \\
 & \textbf{SEER} & \textbf{0.929} & \textbf{0.930} & \multicolumn{1}{r}{\textbf{0.930}} & \textbf{0.930} & \textbf{0.928} & \textbf{0.928} & \textbf{0.928} \\ \hline
\multirow{10}{*}{Twitter} & Att\_RNN~\cite{Att-RNN} & 0.664 & 0.749 & 0.615 & 0.676 & 0.589 & 0.728 & 0.651 \\
 & EANN~\cite{EANN} & 0.648 & 0.810 & 0.498 & 0.617 & 0.584 & 0.759 & 0.660 \\
 & MVAE~\cite{MVAE} & 0.745 & 0.801 & 0.719 & 0.758 & 0.689 & 0.777 & 0.730 \\
 & SpotFake~\cite{Spotfake} & 0.771 & 0.784 & 0.744 & 0.764 & 0.769 & 0.807 & 0.787 \\
 & SAFE~\cite{safe} & 0.762 & 0.831 & 0.724 & 0.774 & 0.695 & 0.811 & 0.748 \\
  & CAFE~\cite{chen2022cross} & 0.806 & 0.807 & 0.799 & 0.803 & 0.805 & 0.813 & 0.809 \\
 & MCAN~\cite{MCAN} & 0.809 & 0.889 & 0.765 & 0.822 & 0.732 & 0.871 & 0.795 \\
 & HMCAN~\cite{HMCAN} & 0,897 & \textbf{0.971} & 0.801 & 0.878 & 0.853 & 0.979 & 0.912 \\
 & CMMTN\cite{CMMTN} & 0.903 & 0.870 & 0.917 & 0.892 & 0.927 & 0.881 & 0.903 \\
 & CCGN~\cite{CCGN} & 0.906 & 0.961 &0.748 & 0.841 & 0.886 &\textbf{0.984} & 0.933 \\
 & \textbf{SEER} & \textbf{0.931} & 0.853 & \textbf{0.983} & \textbf{0.914} & \textbf{0.989} & 0.900 & \textbf{0.942} \\ \hline
\end{tabular}}
\end{center}
\label{tab:all}
\end{table*}

{
\setlength{\belowcaptionskip}{0cm}
\begin{table}[t]
\setlength{\abovecaptionskip}{0cm}
\caption{Ablation studies on different modalities.}
\centering
\resizebox{0.45\textwidth}{!}{
\begin{tabular}{ccccc}
\hline
\multirow{2}{*}{Dataset} & \multirow{2}{*}{Methods} & \multirow{2}{*}{Accuracy} & \multicolumn{2}{c}{F1-score} \\ \cline{4-5} 
 &  &  & Fake News & Real News \\ \hline
\multirow{4}{*}{Weibo} & w/o Text & 0.802 & 0.810 & 0.794 \\
 & w/o Images & 0.903 & 0.902 & 0.905 \\
 & w/o MSEM & 0.914 & 0.914 & 0.914 \\
 & \textbf{SEER} & \textbf{0.929} & \textbf{0.930} & \textbf{0.928} \\ \hline
\multirow{4}{*}{Twitter} & w/o Text & 0.905 & 0.878 & 0.923 \\
 & w/o Images & 0.846 & 0.788 & 0.879 \\
 & w/o MSEM & 0.861 & 0.815 & 0.889 \\
 & \textbf{SEER} & \textbf{0.931} & \textbf{0.914} & \textbf{0.942} \\ \hline
\end{tabular}
}
\label{ablation1}
\end{table}}

\subsection{Experimental Setups}
\noindent{\textbf{Dataset Description.}}
We conduct experiments on two real-world datasets, i.e., Weibo~\cite{Att-RNN} and Twitter~\cite{twitter}.
Here, we mainly focus on text and image information.
Weibo contains 3643 real news and 4203 fake news with 9528 images.
Twitter contains 8720 real news and 7448 fake news with 514 images.
We follow the same preprocessing steps and data split as the benchmark on the two datasets.

\noindent{\textbf{Implementation Details.}}
All experiments are conducted on a single NVIDIA RTX A6000 GPU with 48GB memory. For text and caption, the lengths 
$m$ and $z$ are set to 300.
For image regions, $n$ is set to $49$.
The CLIP model used is ViT-B/16.
The number of heads $m$ is set to 8.
The number of experts $k$ is set to 10.
For Weibo, the parameters $\lambda$, $\alpha$ and $\beta$ are set to 0.75, 0.45, and 0.3, respectively.
For Twitter, the parameters $\lambda$, $\alpha$ and $\beta$ are set to 0.25, 0.65, and 0.3, respectively.
We train the model for 30 epochs by using Adam optimizer with a learning rate of 0.001, and the batch size is 16.
To evaluate performance, we adopt accuracy, precision, recall, and F1-score as metrics.

\noindent{\textbf{Considered Baselines.}}
For comparison, we consider several state-of-the-art baselines:
1) Att\_RNN~\cite{Att-RNN} proposes the RNN with attention mechanism, which integrates textual, visual and social context features;
2) EANN~\cite{EANN} designs the event discriminator and adversarial networks to extract event-invariant features for rumor detection;
3) MVAE~\cite{MVAE} introduces a variational autoencoder to learn the shared features of text and images;
4) SpotFake~\cite{Spotfake} employs the pre-trained BERT and VGG-19 to extract textual and visual features;
5) SAFE~\cite{safe} measures cross-modal similarity between textual and visual features for fake news detection; 
6) CAFE~\cite{chen2022cross} measures cross-modal ambiguity to help aggregate unimodal and multimodal features;
7) MCAN~\cite{MCAN} introduces multiple co-attention layers to fuse features for learning the inter-modal relationships;
8) HMCAN~\cite{HMCAN} proposes a hierarchical multimodal contextual attention network to learn inter- and intra-modality relationships; 
9) CMMTN~\cite{CMMTN} employs a mask-attention mechanism to learn intra- and inter-modal relationships, and incorporates curriculum-driven PU learning to enhance performance;
10) CCGN~\cite{CCGN} improves image-text alignment via contrastive learning and fuses consistency learning with fine-grained cross-modal interaction.

\subsection{Overall Performance}
After extensive experiments, the obtained results are depicted in Table \ref{tab:all}. We find that our proposed SEER significantly outperforms all the considered baselines in terms of Accuracy and F1-score, which demonstrates the effectiveness of our model.
Although Att\_RNN, EANN, and MVAE integrate multimodal features, they rely on simple fusion methods like concatenation without addressing cross-modal semantic gaps, resulting in weak feature fusion and suboptimal detection.
Moreover, they use LSTM, Text-CNN and Bi-LSTM respectively to extract text features, while Spotfake leverages BERT to obtain more powerful representations, achieving better performance and highlighting the importance of pre-trained language models in improving detection.
SAFE and CAFE learn the semantic consistency and similarity between different modalities, demonstrating the effectiveness of improving image-text alignment and learning consistency.
MCAN and HMCAN achieve better performance by introducing the co-attention mechanism, indicating the effectiveness of cross-modal semantic interaction and fine-grained fusion.
CMMTN employs a mask-attention mechanism to capture both intra- and inter-modal relationships, filtering out irrelevant features and minimizing noise, while the incorporation of a curriculum-driven PU learning approach further improves its performance.
CCGN improves image-text alignment via contrastive learning, and achieves superior performance
by integrating consistency learning and GNN for fine-grained cross-modal interactions.

However, our model achieves the best performance.
We attribute the superiority of proposed SEER to two reasons:
1) We use large multimodal models to improve image understanding and cross-modal alignment, designing interaction processes for semantic enhancement of unimodal and multimodal features; 2) We consider the importance of emotions from different modalities and introduce the perceived relationship between news authenticity and emotional tendencies to optimize emotion features for fake news detection.

\subsection{Ablation Study}
\noindent{\textbf{Impact of Different Modalities.}}
We analyze the importance of different modalities by removing text, image, and fusion features, which are denoted as w/o Text, w/o Images, and w/o MSEM, respectively. 
The results are shown in Table \ref{ablation1}.
We find that the absence of the textual, visual, or fusion features degrades performance, which indicates that only utilizing unimodal features is insufficient.
Additionally, text conveys the meaning of news more effectively than images in Weibo. Conversely, images contain richer clues than text in Twitter.
Finally, Weibo exhibits better initial text-image alignment compared to the Twitter, which confirms that the necessity of semantic enhancement and improving text-image alignment.

{
\setlength{\belowcaptionskip}{0cm}
\begin{table}[t]
\setlength{\abovecaptionskip}{0cm}
\caption{Ablation studies on semantic enhancement.}
\centering
\resizebox{0.45\textwidth}{!}{
\begin{tabular}{ccccc}
\hline
\multirow{2}{*}{Dataset} & \multirow{2}{*}{Methods} & \multirow{2}{*}{Accuracy} & \multicolumn{2}{c}{F1-score} \\ \cline{4-5} 
 &  &  & Fake News & Real News \\ \hline
\multirow{5}{*}{Weibo} & w/o CLIP & 0.909 & 0.908 & 0.909 \\
 & w/o Captions & 0.912 & 0.912 & 0.912 \\
 & w/o CA & 0.919 & 0.919 & 0.918 \\
 & w/o SA & 0.924 & 0.924 & 0.923 \\
 & \textbf{SEER} & \textbf{0.929} & \textbf{0.930} & \textbf{0.928} \\ \hline
\multirow{5}{*}{Twitter} & w/o CLIP & 0.885 & 0.853 & 0.906 \\
 & w/o Captions & 0.918 & 0.890 & 0.933 \\
 & w/o CA & 0.899 & 0.870 & 0.917 \\
 & w/o SA & 0.911 & 0.882 & 0.928 \\
 & \textbf{SEER} & \textbf{0.931} & \textbf{0.914} & \textbf{0.942} \\ \hline
\end{tabular}
}
\label{ablation2}
\end{table}}

\noindent{\textbf{Effectiveness of Semantic Enhancement.}}
We analyze the effectiveness of semantic enhancement by removing related components.
The results are shown in Table \ref{ablation2}, where w/o CLIP, w/o Captions, w/o CA, and w/o SA denote the removal of CLIP, captions from BLIP-2, co-attention blocks, and self-attention, respectively. 
We observe that:
1) The w/o CLIP has the most significant impact, indicating that CLIP can effectively enhance text-image alignment and semantic expression;
2) Captions represent the semantics of images, providing a greater enhancement for Weibo with weaker image features, and a lesser enhancement for Twitter with stronger image features;
3) Co-attention blocks significantly improve performance on Twitter, as it has a lower initial text-image alignment compared to Weibo, indicating the necessity and effectiveness of inter-modal interaction and multimodal semantic enhancement; 
4) The w/o SA demonstrates the beneficial role of intra-modal interaction for rumor detection.

{
\setlength{\belowcaptionskip}{0cm}
\begin{table}[t]
\setlength{\abovecaptionskip}{0cm}
\caption{Ablation studies on emotional reasoning.}
\centering
\resizebox{0.45\textwidth}{!}{
\begin{tabular}{ccccc}
\hline
\multirow{2}{*}{Dataset} & \multirow{2}{*}{Methods} & \multirow{2}{*}{Accuracy} & \multicolumn{2}{c}{F1-score} \\ \cline{4-5} 
 &  &  & Fake News & Real News \\ \hline
\multirow{3}{*}{Weibo} & w/o EERM & 0.916 & 0.915 & 0.916 \\
 & SEER\_I & 0.925 & 0.925 & 0.924 \\
 & \textbf{SEER} & \textbf{0.929} & \textbf{0.930} & \textbf{0.928} \\ \hline
\multirow{3}{*}{Twitter} & w/o EERM & 0.915 & 0.889 & 0.931 \\
 & SEER\_I & 0.922 & 0.897 & 0.937 \\
 & \textbf{SEER} & \textbf{0.931} & \textbf{0.914} & \textbf{0.942} \\ \hline
\end{tabular}
}
\label{ablation3}
\end{table}}

\noindent{\textbf{Effectiveness of Emotional Reasoning.}}
In Table \ref{ablation3}, we use the w/o EERM to denote the removal of the expert emotional reasoning module. 
We observe a significant decline in performance, indicating that adjusting contributions of emotions from different modalities and leveraging the perceived relationship between news authenticity and emotional tendencies can effectively improve detection accuracy.
The SEER\_I is a variant that replaces captions with images for emotion analysis. 
It has less impact on Weibo but a significant impact on Twitter, indicating that captions express emotional tendencies better than images in these two datasets.

\subsection{Parameter Analysis}

\begin{figure}[t]
\centering
\scalebox{0.89}
{\includegraphics[width=1\linewidth]{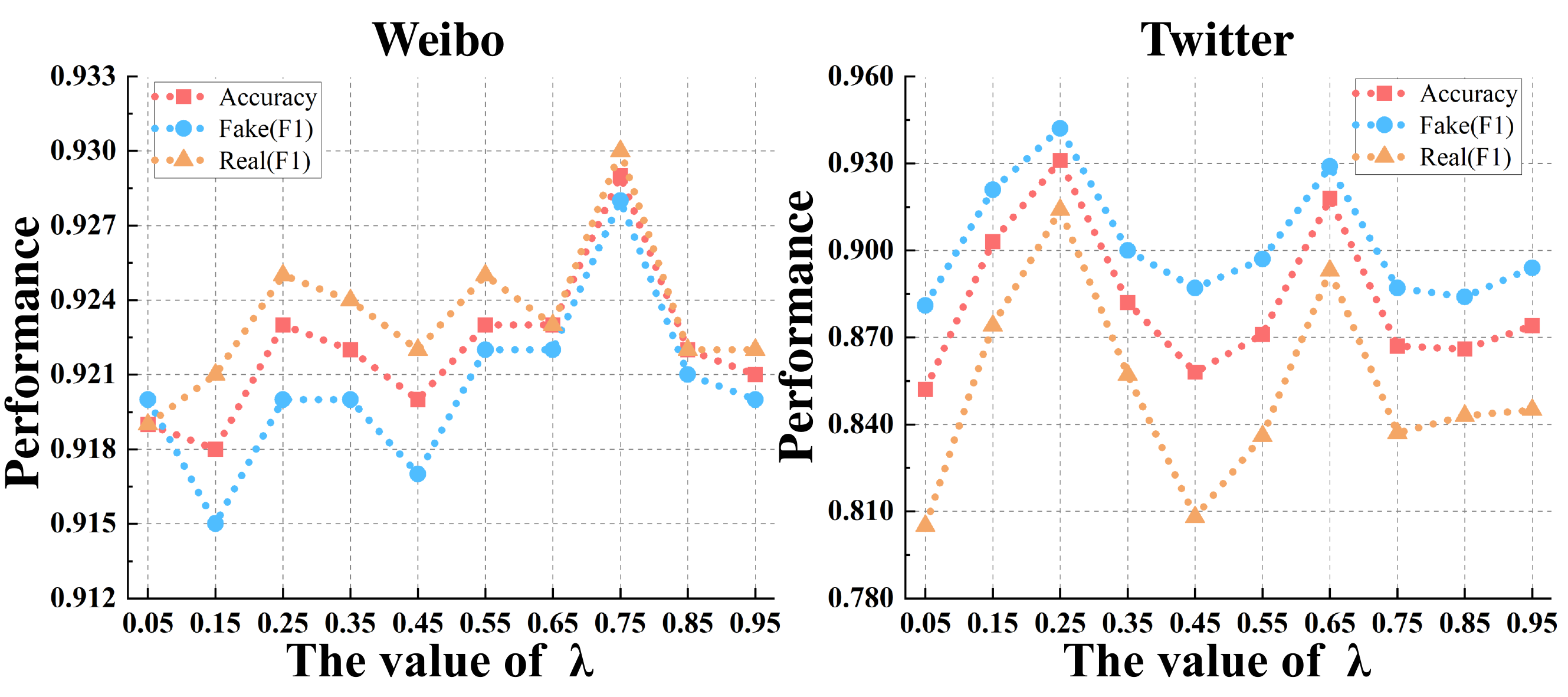}}
\caption{
Impact of emotions from different modalities on two datasets.}
\label{fig:parameter}
\end{figure}

\noindent{\textbf{Impact of Emotions from Different Modalities.}}
We investigate the importance of emotions from different modalities. The results are depicted in Figure~\ref{fig:parameter}, where a larger parameter $\lambda$ represents greater significance of emotions expressed in text.
In Weibo dataset, the overall performance can be regarded as increasing initially and then decreasing. It achieves optimal performance when $\lambda$ is set to 0.75, indicating that the textual content convey richer emotional information compared to the captions.
In Twitter dataset, we can observe the presence of two peaks. Notably, when $\lambda$ is set to 0.25, the model achieves the best performance.
The significant fluctuations indicate notable differences in the expression of emotions between news text and images in Twitter.

\setlength{\abovecaptionskip}{0cm}

\setlength{\abovecaptionskip}{0cm}
\begin{figure}[htb]
\centering
\scalebox{0.55}
{\includegraphics[width=1\linewidth]{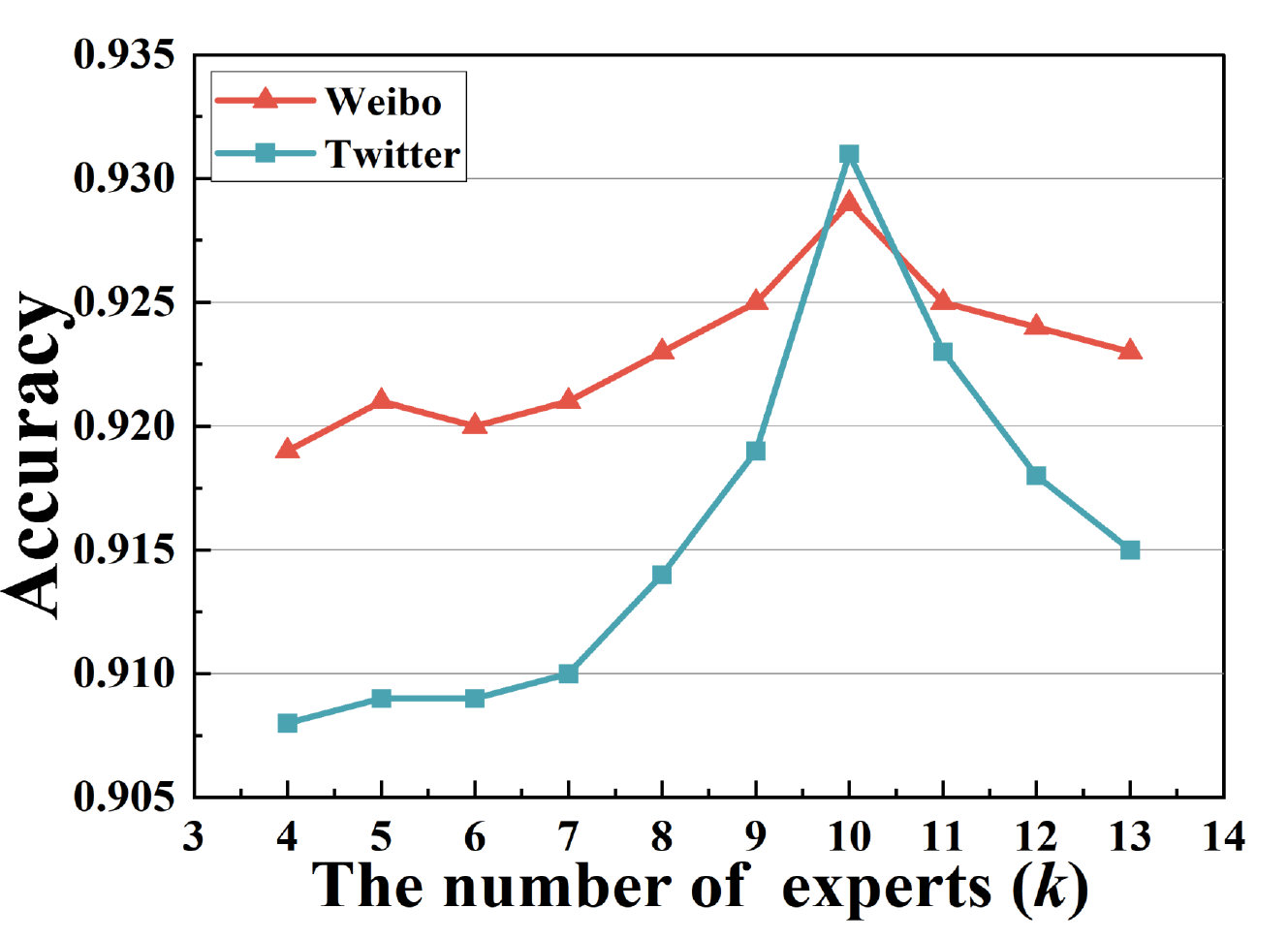}}
\caption{Effect of varying the number of experts ($k$) on classification performance for Weibo and Twitter.}
\label{fig:k}
\end{figure}

\begin{figure}[t]
\centering
\scalebox{0.86}
{\includegraphics[width=1\linewidth]{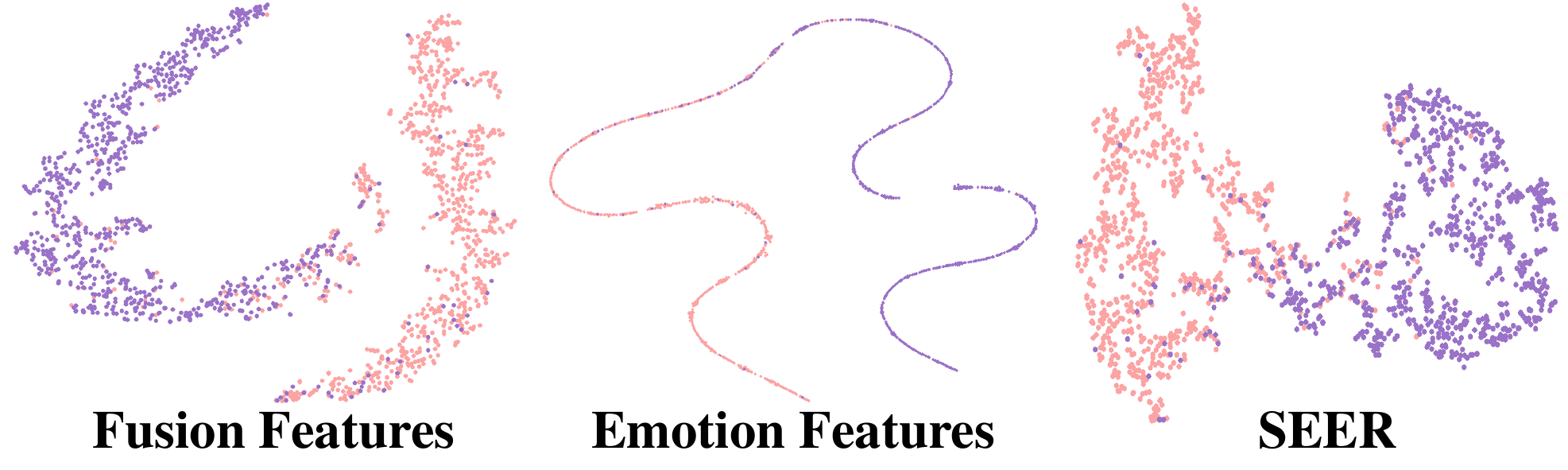}}
\caption{Visualization of the learned representations of test data on Weibo.}
\label{fig:tsne}
\end{figure}

\noindent{\textbf{Impact of the Number of Experts.}}
We conduct experiments with different $k$ on two datasets. 
As revealed in Figure~\ref{fig:k}, the effect of varying $k$ on the detection performance shows a similar trend on both datasets, which can be considered to increase initially and then decrease. This suggests the existence of an optimal $k$.
When $k$ is set to 10, the SEER model achieves the best performance on both datasets.
Additionally, we observe that for Twitter dataset, the impact of varying $k$ on overall performance is much greater than that of Weibo dataset. 
We suppose this phenomenon is due to the stronger correlation between emotions tendencies and news authenticity in Twitter dataset compared to Weibo dataset. Therefore, the effective evaluation and optimization of emotion features are more crucial for Twitter dataset.

\subsection{Visualization of Different Features}

Through t-SNE~\cite{tsne}, we visualize different features learned by SEER on test data of Weibo as in Figure~\ref{fig:tsne}, where the dots of the same color indicate the same label. 
We observe that the extracted fusion features and emotion features are able to clearly distinguish real news from fake ones, demonstrating the effectiveness of multimodal semantic enhancement and expert emotional reasoning.
Finally, the aggregation of all features achieves the best performance, with reduced within-class variance and clear between-class distinction, demonstrating the effectiveness of our SEER in discriminating fake news.

\section{Conclusion}
In this work, we propose a novel \textbf{S}emantic \textbf{E}nhancement and \textbf{E}motional \textbf{R}easoning (SEER) framework to explore the role of news semantic-level enhancement and emotion-level mining for fake news detection.
We generate summarized captions for image semantic understanding and employ products of large multimodal models for semantic enhancement. Furthermore, we consider the varying contributions of emotions from different modalities and utilize the perceived relationship between news authenticity and emotional tendencies to optimize emotion features. 
Experimental results on real-world datasets demonstrate the effectiveness of our SEER.

\bibliographystyle{IEEEtran}
\balance
\bibliography{SEER-SMC2025}

\end{document}